\documentstyle[referee,12pt]{l-aa}

\input{psfig}

\def\Real{{\rm I\mathchoice{\kern-0.70mm}{\kern-0.70mm}{\kern-0.65mm}%
  {\kern-0.50mm}R}}

\font \bolditalics = cmmib10
\def\bx#1{\leavevmode\thinspace\hbox{\vrule\vtop{\vbox{\hrule\kern1pt
        \hbox{\vphantom{\tt/}\thinspace{\bf#1}\thinspace}}
      \kern1pt\hrule}\vrule}\thinspace}

\def \vc #1{{\textfont1=\bolditalics \hbox{$\bf#1$}}}

\def\kg{{\bf k}}

\def\sg{{\bf s}}

\def\varthetag{{\vc \vartheta}}

\begin{document}

%\input{psfig.sty}
%\input{epsf.sty}
%\input{macros.tex}
% \input macros-nofig

%%%%%%%%%%%%%%%%%%%%%%%%%%%%%%
%\centerline{\Large{\bf Scale dependence of the bias investigated by weak lensing}}
%\bigskip
%\centerline{\bf L. van Waerbeke}
%\bigskip
%\noindent
%\centerline{\bf Max-Planck-Institut f\"ur Astrophysik} 
%\centerline{\bf Postfach 1523} 
%\centerline{\bf D-85740 Garching, Germany}
%\hfill\break
%$^2$ Institut d'Astrophysique de Paris, 98 bis boulevard Arago,
%F-75014 Paris, France, \hfill\break
%$^3$ DEMIRM, Observatoire de Paris, 61 Avenue de l'Observatoire,
%F-75014 Paris, France,\hfill\break

%\bigskip
   \thesaurus{12         % A&A Section 12: Cosmology
              (12.04.1;  % Galaxies: evolution,
               12.07.1;  % (Galaxies:) quasars: general:,
               12.12.1)} % Cosmology: miscellaneous.

   \title{Scale dependence of the bias investigated by weak lensing}

   \author{L. van Waerbeke}

   \offprints{waerbeke@mpa-garching.mpg.de}

   \institute{Max-Planck-Institut f\"ur Astrophysik, Karl-Schwarzschild-Str. 1, Postfach 1523, 
D-85740 Garching, Germany}

\date{Received, , Accepted, }

\maketitle

\begin{abstract}

The statistical analysis of the lensing effects coupled with the statistical analysis of
the number counts is a tool to probe directly the relation between the mass and the light.
In particular, some properties of the bias parameter can be
investigated. The correlation between the shear of a given population of galaxies,
and the number counts of a different population of galaxies along the same line of
sight is calculated
for the linear and the non-linear power spectra of density fluctuations
for different cosmologies. The estimator $R$ defined as the ratio of this correlation and the
variance of the number counts is
inversely proportional to the bias parameter. The signal-to-noise ratio of $R$
shows a significant decrease in the non-linear regime where the number of
galaxies per smoothing area is small. At these scales, the noise is dominated by the 
intrinsic ellipticities of the galaxies and by the shot noise, the
cosmic variance playing a minor role. Hence, only galaxy
samples larger than one square degree may allow a precise determination of $R$.
Unfortunately, $R$ is highly dependent on the cosmological model, which makes a direct
measure of the bias quite difficult. However, it is showed that $Rb$ is independent on the power
spectrum and the smoothing scale, thus $R$ is a direct measure of the inverse of the
bias times a function of the cosmological parameters. From $R$, a new
estimator $\cal R$ is defined which is only sensible to the scale dependence of the bias. It
is showed that with a sample of $25$ square degrees, one can measure a scale variation of the bias
larger than 20\% in the $1'$ to $10'$ scale range, almost independently of the cosmological
parameters, the redshift distribution of the galaxies, and the power spectrum, which affect the 
estimate of the variation of $b$ from $\cal R$ by less than 2\%.

\end{abstract}

%\vfill\eject

\section{Introduction}

The statistical analysis of the weak lensing effects can be used to probe the projected
mass distribution in the Universe and to constrain the cosmological parameters.
The variance of the gravitational shear can be used (Blandford et al. 1991, Kaiser 1992,
Villumsen 1996a and Kaiser 1996), and both the variance and the skewness of the convergence can
probe, almost independently, the shape of the power spectrum and the cosmological parameters
(Bernardeau et al. 1997). These papers
provide analytical and numerical estimates of these statistical estimators
using the dominant order of the perturbation theory of the large-scale structure
formation. Hence, these calculations are limited to scales larger than a few Megaparsecs\footnote{
Which corresponds to angular scales larger than $10'$ at a redshift of $0.4$.}
To take into account the fully non-linear evolution of the density contrast,
Jain \& Seljak (1997) calculated the variance of the shear using the non-linear evolution
of the power spectrum derived by Peacock \& Dodds (1996).
They have shown that the signal is increased by a factor two to three, compared to the
linear perturbation theory. More recently, Schneider et al. (1997) (hereafter SvWJK)
calculated the variance and the skewness of the aperture mass $M_{\rm ap}$, which is defined as the
convergence measured with a compensated filter.
This statistic, inspired by the cluster lensing analysis (Kaiser et al. 1994, and
Schneider 1996), has the
nice property to be directly measurable from the shape of galaxies, while this is not the case
for the convergence alone which is not observable.

The statistical analysis of the lensing effects may also be fruitfully coupled with the statistical
analysis of the galaxy number counts. Villumsen (1996b) has shown how the 2-point correlation
function of very distant galaxies is changed due to lensing by the foreground
structures. Basically, this magnification bias effect produces an enhancement of the correlation
at small scales. Moessner \& Jain (1997) extended these calculations into the non-linear power
spectrum.
Since the bias parameter enters explicitly in the calculations, this kind of study provides
an original way to analyse the bias properties. However, they have shown that the lensing effect
remains very small, even in the Hubble Deep Field (Villumsen et al. 1996).
Sanz et al. (1997)
made the first systematic analysis of the shear-number counts cross-correlations
for different cosmological models. They have shown that, in the non-linear regime,
 the signal is significantly enhanced.

In a very recent paper, Schneider (1997) calculated the correlation $\langle M_{\rm ap}{\cal N}
\rangle$ between $M_{\rm ap}$ and the galaxy number counts $\cal N$
filtered by a compensated filter. Using linear perturbation
theory, he has shown that $\langle M_{\rm ap}{\cal N}\rangle $ is much easier to measure than
$\langle M_{\rm ap}^2\rangle $ because the corresponding signal-to-noise ratio is increased by the correlation. 
Using this statistic, a 
signal-to-noise ratio per field of up to $0.5$ is achieved, whatever the scale, while it
is close to $0.1$ for $\langle M_{\rm ap}^2\rangle $ at $10'$.

The aim of this paper is to focus on some properties of this aperture mass-number counts
correlation at small scales, and its capability to measure the scale dependence of the
bias parameter for a given redshift distribution of the background sources and the foreground
 galaxies.
The $\langle M_{\rm ap}{\cal N}\rangle $ statistic as calculated in Schneider (1997) will be used, but the
calculations will be extended into the non-linear regime to the scale of $1'$, for a variety of
cosmological models. From
an estimate of the signal-to-noise ratio, it is shown how the scale dependence of the bias
parameter may be measured with reasonable accuracy for a wide range of scales ($1'$ to $10'$
for a survey of $25$ square degrees), irrespective of the assumed cosmological model. This provides
a way to probe how the bias varies with scale.

Section 2 presents a summary of the $M_{\rm ap}$, $\cal N$ and $M_{\rm ap}\cal N$
statistics. A new compensated filter is introduced, which permits simple analytical calculations
in real and Fourier space.
In Section 3, the correlation $\langle M_{\rm ap}{\cal N}\rangle $ is calculated in the non-linear
regime, as well as its signal-to-noise ratio per field. In Section 4, a new statistical
estimator is defined, which is able to probe the scale dependence of the bias.
The measurability of this estimator
is discussed in the light of the previous signal-to-noise analysis.

\section{Statistics with compensated filters}

\begin{figure}
\centerline{
\psfig{figure=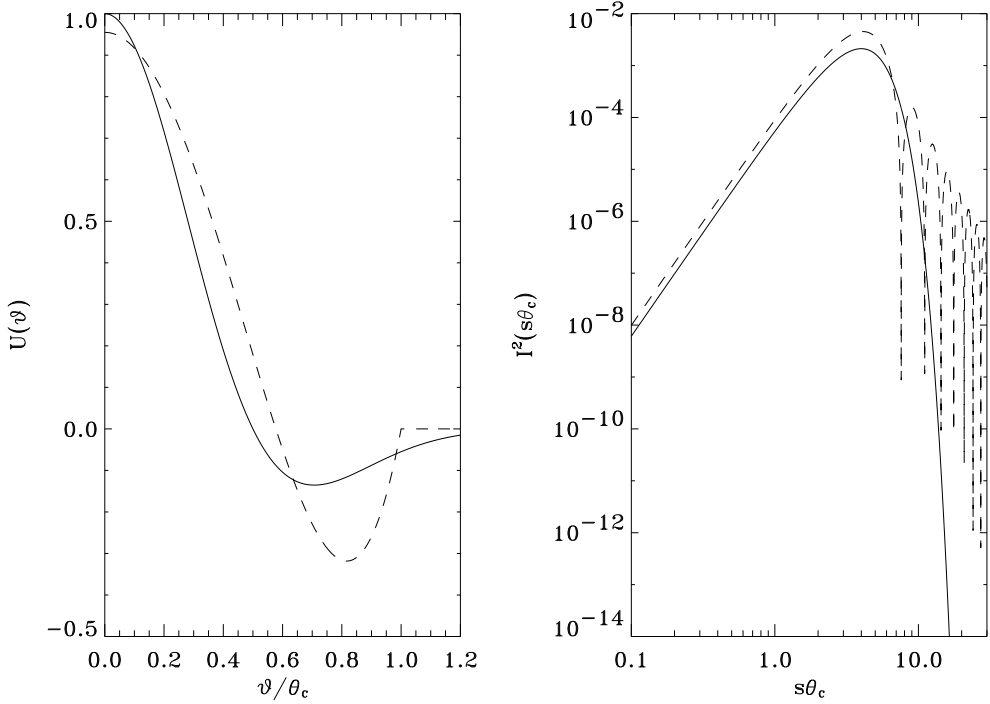,width=14cm}}
\caption{\label{plot_filtre.ps} On the left, the filters (\ref{filt_1}) (dashed line, for $l=1$)
and (\ref{U_theta_c}) (solid line). On the right, their 2-D Fourier transforms, as
defined in the text.}
\end{figure}

In SvWJK, the aperture mass $M_{\rm ap}$ within a circular radius $\theta_{\rm c}$ is defined,

\begin{equation}
M_{\rm ap}(\theta_{\rm c})=\int_0^{\infty} {\rm d}^2\vartheta U(\vartheta) \kappa(\varthetag),
\end{equation}
where $\kappa(\varthetag)$ is the convergence at the angular position $\varthetag$, and
$U(\vartheta)$ is the compensated filter,

\begin{equation}
U(\vartheta)={(l+2)^2\over \pi \theta_{\rm c}^2}\left[1-\left({\vartheta\over \theta_{\rm c}}\right)^2\right]^l\left[
{1\over l+2}-\left({\vartheta\over \theta_{\rm c}}\right)^2\right],
\label{filt_1}
\end{equation}
and $U(\vartheta)=0$ if $\vartheta>\theta_{\rm c}$. $l$ is an integer parameter, and 
$\theta_{\rm c}$ is
the scale of smoothing.  The aperture mass is related to the tangential part of the observable
shear $\gamma_{\rm t}$,

\begin{equation}
M_{\rm ap}(\theta_{\rm c})=\int_0^{\infty} {\rm d}^2\vartheta Q(\vartheta) \gamma_{\rm t}(\varthetag),
\end{equation}
where $Q(\vartheta)={2\over \vartheta^2}\int_0^\vartheta d\vartheta' \vartheta'U(\vartheta')-
U(\vartheta)$ (Kaiser et al. 1994). The compensated filters cut-out the power at the scale $\theta_{\rm c}$ , which leads to
narrow filters in the Fourier space, well-localized around a particular
frequency. However, in SvWJK, the filters $U(\vartheta)$ and
$Q(\vartheta)$ are defined on a compact space ($U(\vartheta)=Q(\vartheta)=0$ if 
$\vartheta>\theta_{\rm c}$), which produces oscillations in Fourier space,
and leads to moderately difficult analytical and numerical calculations. A new compensated filter is
introduced here, with a non-compact definition range, but with a sufficiently fast decrease
such that it can be considered as a compact filter for the practical use,

\begin{equation}
U(\vartheta)={1\over {\theta_{\rm c}}^2} \left(1-4{\vartheta^2\over {\theta_{\rm c}}^2}\right)~\exp{\left( -4{\vartheta^2\over {\theta_{\rm c}}^2}\right)},
\label{U_theta_c}
\end{equation}
and $Q(\vartheta)={4\over {\theta_{\rm c}}^2} {\vartheta^2\over {\theta_{\rm c}}^2}~
\exp{\left(-4{\vartheta^2\over {\theta_{\rm c}}^2}\right)}$. The 2-D Fourier transform of $U$ is,

\begin{equation}
{1\over 2\pi} \int{\rm d}^2\vartheta~U(\vartheta)~e^{{\rm i}\sg\cdot\varthetag}={1\over 128}s^2\theta_{\rm c}^2~
\exp{\left( -{s^2\theta_{\rm c}^2\over 16}\right)}={\rm :}~I\left({s\theta_{\rm c}\over 4}\right).
\end{equation}
Thus  $I(\eta)={1\over 8}
\eta^2~e^{-\eta^2}$ which peaks at $\eta=1$. In Fourier space,
both filters
(\ref{filt_1}) and (\ref{U_theta_c}) peak at $s\simeq 4/\theta_{\rm c}$ and they have the same width.
Figure \ref{plot_filtre.ps} displays these two filters and the square of their 2-D Fourier transform. 

The filter $U$ defined as (\ref{U_theta_c}) belongs to the family of wavelets. Note that it
would be ideal
to use wavelets, which are much more localized in the Fourier space, to probe accurately the power
spectrum of the projected mass distribution. Unfortunately, for our particular case, the real space
filter of most of the wavelets oscillates too rapidly to be sampled enough by the discrete distribution
of the galaxies.
The filter (\ref{U_theta_c}) seems to be a good compromise, and it is used for this work.

In SvWJK, the dispersion of $M_{\rm ap}(\theta_{\rm c})$ is calculated to be

\begin{equation}
\langle M_{\rm ap}^2(\theta_{\rm c})\rangle ={9\pi \over 2} \left(H_0\over c\right)^4 \Omega^2\int_0^{w_H}~{\rm d}w~{g^2(w)\over 
a^2(w)} \int~{\rm d}s~s~P\left({s\over f_K(w)},w\right)~I^2(s\theta_{\rm c}),
\label{map_def}
\end{equation}
where $w(z)$ is the comoving distance to a redshift $z$ (and $w_H=w(\infty$)), $f_K$ is the comoving
angular diameter
distance, and $P$ is the time-evolving 3-D power spectrum. $\Omega$ is the density parameter, and
$a$ is the cosmic expansion factor. The function $g(w)=\int_w^{w_H}~{\rm d}w'~p_{\rm b}(w')~f_K(w'-w)/f_K(w')$
depends on the redshift distribution of the sources $p_{\rm b}(w)~{\rm d}w=\tilde p_{\rm b}(z)~
{\rm d}z$.

The aperture mass $M_{\rm ap}$ should now be correlated with a distribution of galaxies along
the same line of sight. From a practical point of view, it is easier to measure
the redshift of the nearest galaxies, and in the following we assumed that the number counts
are done in a foreground distribution of galaxies, but it is worth to note that in general,
this restriction is not necessary.
Following the standard bias theory, the galaxy density contrast $\delta_{\rm g}$ is related to
the mass density contrast $\delta$,

\begin{equation}
\delta_{\rm g}=b~\delta.
\end{equation}
The expected number density contrast of galaxies in the direction $\varthetag$ is then

\begin{equation}
\Delta n_g(\varthetag)={N(\varthetag)-\bar N\over \bar N}=b\int~{\rm d}w~p_{\rm f}(w)~\delta(f_K(w)\varthetag,w),
\end{equation}
where $\bar N$ is the mean number density of galaxies, $N(\varthetag)$ the number density of galaxies
on the direction $\vartheta$, and $p_{\rm f}(w)~{\rm d}w=\tilde p_{\rm f}(z)~{\rm d}z$ is the
redshift distribution of the foreground galaxies. If the filtered number counts are defined as
${\cal N}(\theta_{\rm c})=\int {\rm d}^2\vartheta~U(\vartheta) \Delta n_g(\varthetag)$,
the dispersion of the galaxy number counts is given by (Schneider 1997)

\begin{equation}
\langle {\cal N}^2(\theta_{\rm c})\rangle =2\pi b^2 \int {\rm d}w {p_{\rm f}^2(w)\over f_K^2(w)} \int {\rm d}s~s~P\left({s\over
f_K(w)},w\right) I^2(s\theta_{\rm c}),
\label{N2_def}
\end{equation}
while the cross-correlation $\langle M_{\rm ap}(\theta_{\rm c}){\cal N}(\theta_{\rm c})\rangle $ is

\begin{equation}
\langle M_{\rm ap}(\theta_{\rm c}){\cal N}(\theta_{\rm c})\rangle =3\pi \left({H_0\over c}\right)^2 \Omega b \int {\rm d}w
{p_{\rm f}(w) g(w)\over a(w) f_K(w)}\int {\rm d}s~s P\left({s\over f_K(w)},w\right) I^2(s\theta_{\rm c}).
\label{MN_def}
\end{equation}

\section{Numerical estimates}

\begin{figure}
\centerline{
\psfig{figure=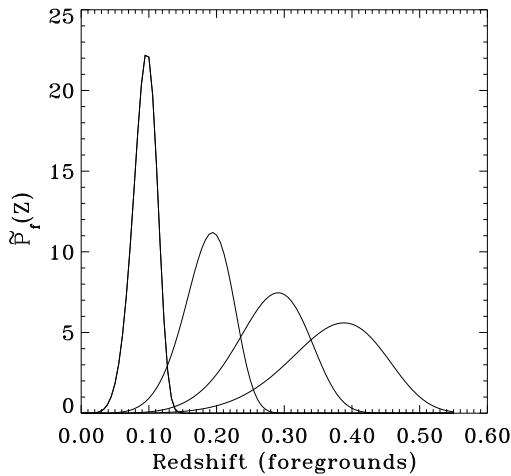,width=9cm}}
\caption{\label{plot_red.ps} The four redshift distributions used for the foreground galaxies,
following the distribution (\ref{p_z_def}). The widths are larger than the characteristic
error of the photometric redshift determination, which is roughly $0.05$ (Pell\'o et al. 1996).}
\end{figure}

Now, numerical estimates of Eq.(\ref{MN_def}) are given. Four models with
a CDM-like spectrum 
given in Bardeen et al. (1986) have been choosen. Three models are normalized to unity ($\sigma_8=1$), with a
shape parameter $\Gamma=0.25$. The corresponding cosmological parameters are
EdS, open ($\Omega=0.3,\Lambda=0$) and flat ($\Omega=0.3, \Lambda=0.7$) Universes.
The fourth model is an EdS Universe with $\Gamma=0.5$ and $\sigma_8=0.6$, corresponding to
the cluster abundance normalization.
The background and foreground populations of galaxies are assumed to follow the normalized redshift
distribution

\begin{equation}
\tilde p_i(z)=\Gamma^{-1}\left({1+\alpha\over \beta}\right){\beta \over z_i}\left({z\over z_i}\right)^\alpha~\exp
\left[-\left({z\over z_i}\right)^\beta\right],
\label{p_z_def}
\end{equation}
where $z_i$ is $z_{\rm b}$ or $z_{\rm f}$, depending on the galaxy population considered.
For the background sources we assume $\alpha=2,~\beta=1.5,~z_{\rm b}=1$, and for the foreground
we take $\alpha=5,~\beta=6$, with four different values for $z_{\rm f}=(0.1,~0.2,~0.3,~0.4)$. The foreground
distribution
is quite narrow, and does not overlap much with the background galaxies. We assume that it is
no problem to derive $\tilde p_{\rm f}(z)$ from observations since the redshifts $z_{\rm f}$ are
small, making these
distribution observable directly from photometric redshifts, if spectroscopic redshifts are not
available. Figure 
\ref{plot_red.ps} displays the four redshift distributions used for the foreground galaxies
corresponding to the four choices of $z_{\rm f}$.

\begin{figure}
\centerline{
\psfig{figure=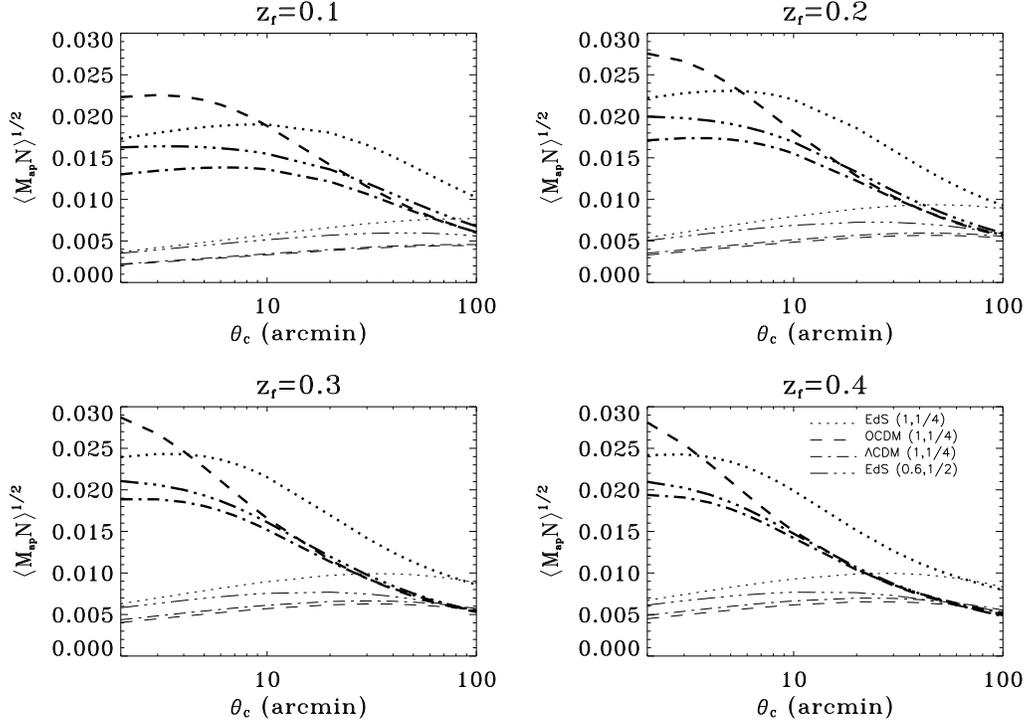,width=14cm}}
\caption{\label{plot_MN.ps} The values of $\langle M_{\rm ap}(\theta_{\rm c}){\cal N}(\theta_{\rm c})\rangle ^{1/2}$ versus the
smoothing scale are shown. The case of a non-linear power spectrum is plotted in thick lines using the
Peacock \& Dodds formula. The thin lines show the linear power spectrum. Different types of
lines correspond to different cosmologies. The top-left, top-right, bottom-left, bottom-right
plots correspond to the foreground redshift distribution according to (\ref{p_z_def}) with
$z_{\rm f}=(0.1,~0.2,~0.3,~0.4)$. For the background galaxies we choose $z_{\rm b}=1$. The
cosmological
models are $\Omega=1,~\Lambda=0,~\sigma_8=1,~\Gamma=0.25$ (EdS (1,1/4)), 
$\Omega=0.3,~\Lambda=0,~\sigma_8=1,~\Gamma=0.25$ (OCDM (1,1/4)),
$\Omega=0.3,~\Lambda=0.7,~\sigma_8=1,~\Gamma=0.25$ ($\Lambda$CDM (1,1/4)),
$\Omega=1,~\Lambda=0,~\sigma_8=0.6,~\Gamma=0.5$ (EdS (0.6,1/2))}
\end{figure}

In Figure \ref{plot_MN.ps}, $\langle M_{\rm ap}(\theta_{\rm c}){\cal N}(\theta_{\rm c})
\rangle ^{1/2}$ versus
smoothing angle $\theta_{\rm c}$ is plotted for the four cosmological models, for both the linear
(thin curves) and the
non-linear (thick curves) power spectra. The four plots correspond to the four different foreground
redshift distributions. At small scales and for the non-linear power spectrum, the signal is strongly 
enhanced, by more than a factor three. This confirms that the linear 
evolution is negligible at these scales (Sanz et al. 1997).
In Fourier space, the compensated filter peaks around $4/\theta_{\rm c}$
rather than $1/\theta_{\rm c}$ for a top hat filter. Thus, the scale at which the linear and the
non-linear evolution of the structures are equal is shifted to larger scale, compare to the case
of a top-hat filter (Jain \& Seljak 1997).
The curves are shifted towards smaller scales
when the foreground galaxies are located at higher redshifts. This is due to the fact that at higher redshift,
a given angular scale corresponds to a larger physical scale, where the non-linear power spectrum
is closer to the linear power spectrum, and that at high redshift, the power spectrum
is less evolved than at small redshift.
For the non-linear power spectrum , the peak in $\langle M_{\rm ap}(\theta_{\rm c}){\cal N}
(\theta_{\rm c})\rangle ^{1/2}$ is shifted to small scales compared to the linear
power spectrum because of the transfer of power from the larger to the smaller scales (this was
also observed for the $M_{\rm ap}$ statistic, see SvWJK).

\begin{figure}
\centerline{
\psfig{figure=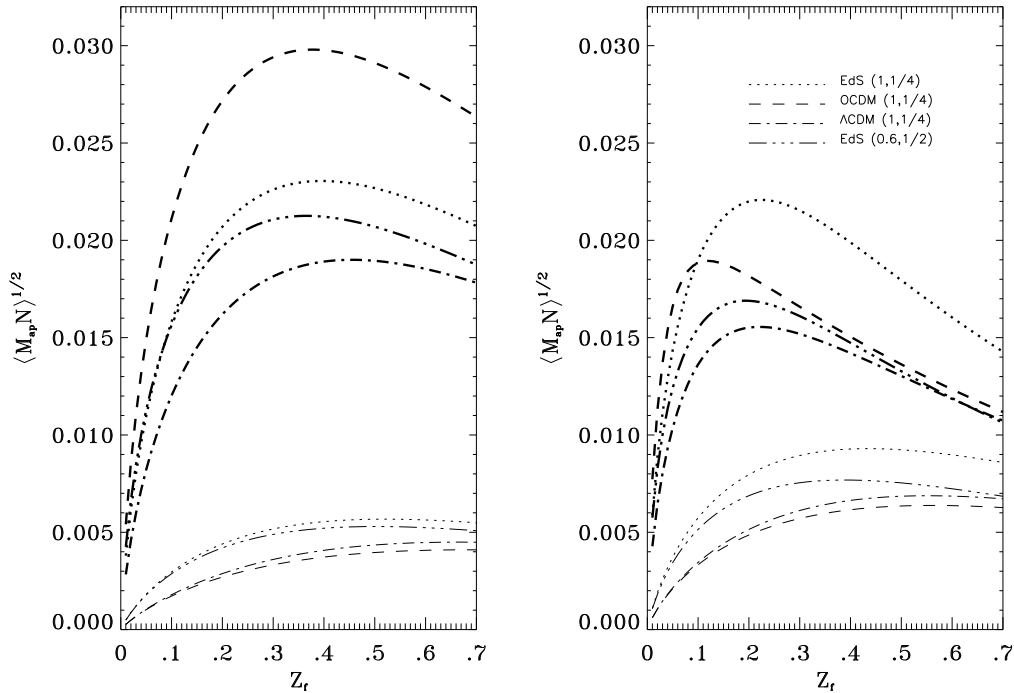,width=14cm}}
\caption{\label{plot_Z_f.ps} The values of $\langle M_{\rm ap}(\theta_{\rm c}){\cal N}(\theta_{\rm c})\rangle ^{1/2}$ versus the
redshift of the foreground galaxies $z_{\rm f}$ given by (\ref{p_z_def}) for two different smoothing
scales ($\theta_{\rm c}=1'$ on the
left and $\theta_{\rm c}=10'$ on the right). The thick lines correspond to the non-linear power
spectrum, the thin ones to the linear power spectrum.}
\end{figure}

Figure \ref{plot_Z_f.ps} shows $\langle M_{\rm ap}(\theta_{\rm c}){\cal N}(\theta_{\rm c})\rangle ^{1/2}$
versus the redshift $z_{\rm f}$ of the foreground galaxies, where the redshift distribution
(\ref{p_z_def}) is
used. The two plots correspond to $\theta_{\rm c}=1'$ (left) and $\theta_{\rm c}=10'$ (right).
At scale of $1'$ the contribution of the linear evolution
to the correlation is again very small. The maximum of
the correlation is obtained for foreground galaxies located at relatively low redshift, compared
to the mean redshift of the background galaxies ($z_{\rm b}=1$). The reason is that for
a given smoothing scale, a
lower redshift corresponds to a higher density contrast, which means a stronger signal. 
Moreover this effect is more important for the non-linear power spectrum where a {\it bump} is clearly
seen compared to the linear power spectrum.

The same practical estimators for $M_{\rm ap}$ and $N$ as those introduced in Schneider (1997)
are used
to estimate the noise of this statistic,

\begin{eqnarray}
\tilde M_{\rm ap}&=&{\pi \theta_{\rm lim}^2\over N_{\rm b}} {\displaystyle \sum_{i=1}^{N_{\rm b}}}~Q(|\vartheta_i|)
\epsilon_{ti}\cr
\tilde N&=&{1\over \bar N}{\displaystyle \sum_{i=1}^{N_{\rm f}}}~U(|\vartheta_i|),
\end{eqnarray}
where $\varthetag_i$ and $\epsilon_{ti}$ are the position and the tangential component of the
ellipticity of the $i^{th}$ galaxy, $\theta_{\rm lim}$ is a radius cut of the filter (otherwise,
it extends to infinity), and
$N_{\rm f}$ and $N_{\rm b}$ are respectively the number of galaxies found in
the foreground population, and those used for the determination of the shear, in the region
limited by the radial cut. $\theta_{\rm lim}$ may be choose arbitrarly, according to 
Figure \ref{plot_filtre.ps} one could decide $\theta_{\rm lim}> 2~\theta_{\rm c}$ in the center
of the whole field, and $\theta_{\rm lim}\simeq 1.2~\theta_{\rm c}$ at the edge of the field.
The dispersion of $\langle M_{\rm ap}(\theta_c){\cal N}(\theta_c)\rangle $ is given by the square root of
the expectation value $E\left[M_{\rm ap}^2(\theta_{\rm c}){\cal N}^2(\theta_{\rm c})\right]-
E^2\left[M_{\rm ap}(\theta_{\rm c}){\cal N}(\theta_{\rm c})\right]$. The signal-to-noise is defined
as the measured signal divided by the standard deviation of $\langle M_{\rm ap}(\theta_{\rm c}){\cal N}
(\theta_{\rm c})\rangle $ obtained in the case of no correlation between $M_{\rm ap}$ and $\cal N$.
This standard deviation is simply the square root of
$E\left[M_{\rm ap}^2(\theta_{\rm c})\right]E\left[{\cal N}^2(\theta_{\rm c})\right]$.
The signal-to-noise ${\cal S}_{\theta_{\rm c}}$
of the cross-correlation $\langle M_{\rm ap}(\theta_{\rm c}){\cal N}
(\theta_{\rm c})\rangle $ for one field is thus (cf Schneider 1997),

\begin{equation}
{\cal S}_{\theta_{\rm c}}={\langle M_{\rm ap}(\theta_{\rm c}){\cal N}(\theta_{\rm c})\rangle \over \left[\langle M_{\rm ap}^2(\theta_{\rm c})\rangle +{G \sigma_\epsilon^2
\over 2 N_{\rm b}}\right]^{1/2}\left[\langle {\cal N}^2(\theta_{\rm c})\rangle +{\tilde G \over 
N_{\rm f}}\right]^{1/2}},
\label{S_N_def}
\end{equation}
where $G=\pi \theta_{\rm lim}^2\int {\rm d}^2\vartheta Q^2(\vartheta)=0.6{\displaystyle \left({
\theta_{\rm lim}\over \theta_c}\right)^2}$, $\tilde G=\pi \theta_{\rm lim}^2\int 
{\rm d}^2\vartheta U^2(\vartheta)=G$, and $\sigma_\epsilon$ is the dispersion of the intrinsic
ellipticities of the galaxies.

\begin{figure}
\centerline{
\psfig{figure=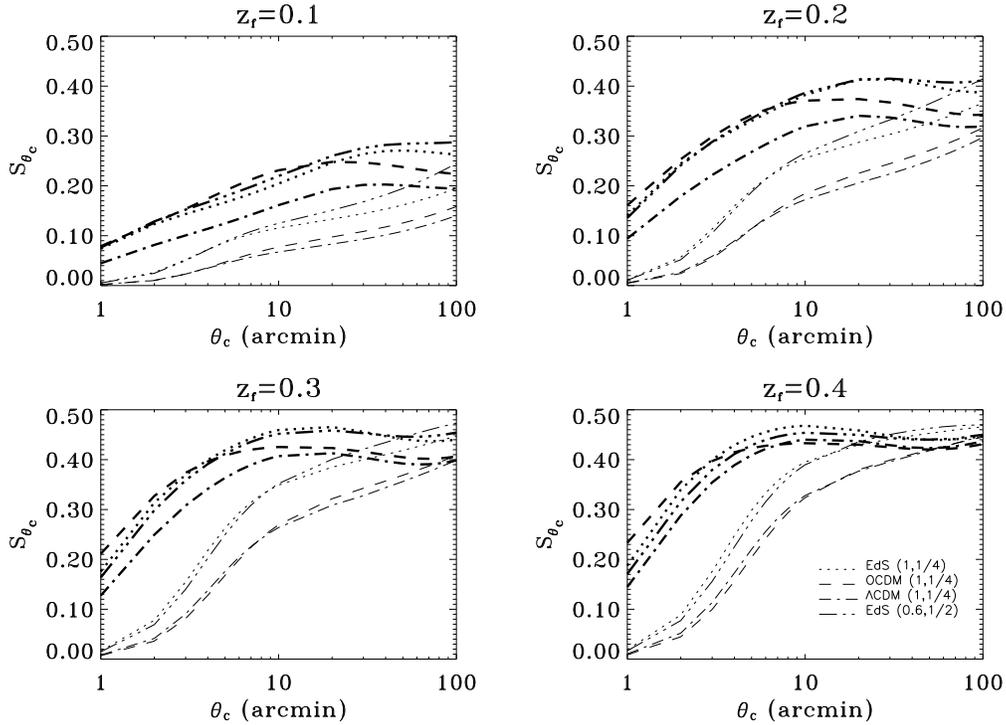,width=14cm}}
\caption{\label{plot_S_N_NL.ps} Signal-to-noise ratio ${\cal S}_{\theta_{\rm c}}$ 
of $\langle M_{\rm ap}(\theta_{\rm c}){\cal N}(\theta_{\rm c})\rangle $ as
defined in the text, versus the smoothing scale. These plots show the expected signal-to-noise for one
field of characteristic radius $\theta_{\rm c}$. The thick lines correspond to the non-linear power
spectrum, the thin ones to the linear power spectrum. At large scale
${\cal S}_{\theta_{\rm c}}$ obtained from the linear power spectrum can be
larger than ${\cal S}_{\theta_{\rm c}}$ of the non-linear power spectrum.}
\end{figure}

Figure \ref{plot_S_N_NL.ps} shows the signal-to-noise ratio given by (\ref{S_N_def}). It is
assumed\footnote{This roughly corresponds to 4 hours exposure on a 4 meter telescope
in the B band (Tyson 1988).}
that the mean number density of the foreground galaxies is
$\bar N\simeq 5~{\rm gal}/{\rm arcmin}^2$ and
that the number density of the background galaxies is $n_{\rm b}=60~{\rm gal}/{\rm arcmin}^2$, with
$\sigma_\epsilon=0.2$. 
We find that ${\cal S}_{\theta_{\rm c}}$ remains almost constant for scales $\theta_{\rm c}>10'$ whatever the cosmology.
This result was found by Schneider (1997) for an EdS model and a power law spectrum. Unfortunately,
this is no longer valid at small scales (few arcmin), where
the effect of intrinsic ellipticities and the discrete distribution of the galaxies become dominant.
The increase of the signal compared to the noise is stronger in the case of the non-linear power
spectrum, this is why at small scales, ${\cal S}_{\theta_{\rm c}}$ is always higher for a non-linear power spectrum.
The redshift of the foreground galaxies is also an important factor, we see that
${\cal S}_{\theta_{\rm c}}$ decreases significantly if $z_{\rm f}=0.1$. Fortunately, even
with this dramatic decrease
of ${\cal S}_{\theta_{\rm c}}$ at small scales, it remains higher than  for
the $\langle M_{\rm ap}^2\rangle $ statistic, where ${\cal S}_{\theta_{\rm c}}$ is close to $0.1$ (SvWJK). As quoted
in Schneider (1997), this makes the
background-foreground correlation statistic more interesting for the detection of cosmic shear,
in particular if the size of the
catalogues used for lensing analysis are not larger than a few square degrees.

\section{Investigation of some bias properties}

Equation (\ref{MN_def}) provides that a direct estimate of the bias parameter, provided
the cosmological model is known. To investigate the bias properties, we define the ratio $R$

\begin{equation}
R={\langle M_{\rm ap}(\theta_{\rm c}){\cal N}(\theta_{\rm c})\rangle \over \langle {\cal N}^2(\theta_{\rm c})\rangle },
\label{R_first}
\end{equation}
which is independent of the normalization of the power spectrum. It is easy to show from 
(\ref{MN_def}) and (\ref{N2_def}) that in the
case of a power law power spectrum $P(k)\propto k^n$, $R$ is independent of scale,

\begin{equation}
R={3\over 2}\left({H_0\over c}\right)^2 {\Omega\over b} {\int~{\rm d}w{p_{\rm f}(w) g(w)D^2_+(w)\over a(w) 
f_K^{(1-n)}(w)}\over \int~{\rm d}w {p_{\rm f}^2(w)D^2_+(w)\over f_K^{(2-n)}(w)}},
\label{R_def}
\end{equation}
where $D_+(w)$ is the linear growth factor of the density perturbations.
This property offers the possibility to measure the scale dependence of the bias (if it exists)
because $R\propto 1/b$, and $R$ is a measurable quantity. A
measure of the bias itself is also possible, but it requires the knowledge of the cosmological
parameters, and the slope of the power spectrum. Moreover, the background and foreground redshift
distributions have to be known. Thus only the effects of a scale dependent bias will be investigated
here.

The true power spectrum is certainly not a power law and
$R$ may depend in a complicated way on the cosmology, the smoothing scale, the power spectrum, and
the redshift distributions. All these things coupled together, $Rb$ is no longer
scale-independent. Fortunately, the compensated filter
$U$ is very narrow in Fourier space, and even for a general power spectrum, it is fully justified
to approximate it locally
as a power law, with a local effective slope $n_{\rm eff}$. Furthermore, since in
practice one can select the foreground galaxies in a narrow redshift range,
both the redshift and wavevector integrations in (\ref{MN_def}) are
very localized. Hence, the local approximation of a general power spectrum by a power law is
very reasonable.

If $b$ is only redshift dependent, this dependence can be absorbed by the function $p_{\rm f}(w)$,
and $R$ remains scale-independent.
However, if $b$ is in addition scale dependent, it means that $b$ depends on the 3-D wavevector $\kg$.
For simplicity let us assume that the redshift and the wavevector dependences are separable.
Fry and Gazta$\tilde{\rm n}$aga (1993) proposed to model such a bias as a convolution,

\begin{equation}
\tilde\delta_{\rm g}(\kg)=\tilde b(\kg)\tilde\delta(\kg).
\end{equation}
In that case, since $b$ is not a constant, $R$ is no longer inversely proportional
to $b$, because it should be included in the projection effects.
This drawback is avoided if
the foreground redshift distribution is narrow, and if the Fourier transform of the aperture
filter is narrow, such that the $s$ integration in (\ref{MN_def}) is performed on a very narrow range,
and $b$ can be approximated as "locally" independent of
redshift and scale. This is the reason for choosing the foreground redshift distributions
(Figure \ref{plot_red.ps}) to be so narrow.

\begin{figure}
\centerline{
\psfig{figure=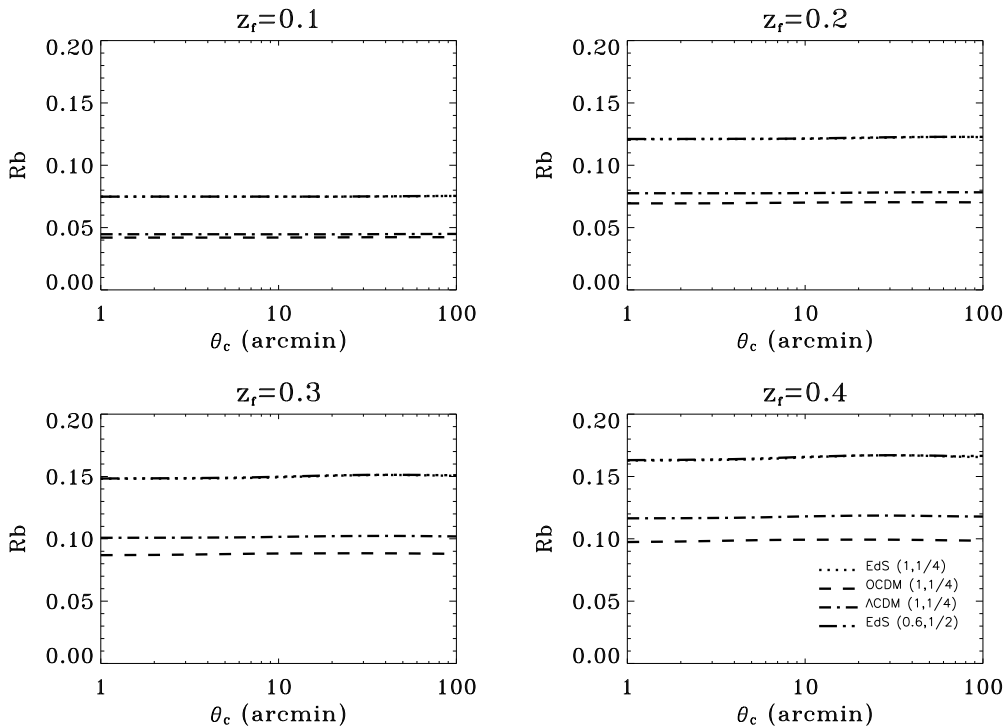,width=14cm}}
\caption{\label{plot_R_val.ps} The values of $Rb$ (Eq. (\ref{R_def})) versus the
smoothing scale are shown. The non-linear and the linear power spectrum
give almost the same results, thus only the
thick lines are plotted. As can be seen, the two EdS cosmological models which differ by the
normalization
and by the shape parameter give also the same results because $R$ does not depend on $\sigma_8$.}
\end{figure}

These remarks are rather phenomenological, but they illustrate what happens in the case of a general
power spectrum for which $R$ is intrinsically scale dependent.
Figure \ref{plot_R_val.ps} shows $Rb$ versus the smoothing scale $\theta_{\rm c}$ for
the cosmological models considered before. The linear and non-linear
power spectra give almost the same results, thus only the thick lines corresponding to the 
non-linear power spectrum are shown here. The
flatness of the curves is remarkable, and confirms that even for a general power spectrum, $R$
depends weakly on scale. The amplitude depends on the cosmology, the shape
of the power spectrum, as well as on the redshift distribution of the foreground and background
galaxies. It is important to note that the non-linear
power spectrum may also be approximated locally as a power law, because
$R$ is independent on scale. This impressive flatness of $R$ versus the scale for
a general power spectrum should be discussed, because it is not so intuitive. Indeed, for non
power-law power spectra, the effective index $n_{\rm eff}$ must be
scale dependent. The very small variation of $R$ with respect to $\theta_{\rm c}$
in Figure \ref{plot_R_val.ps} means
that the value of the effective index has no strong influence on the value of $R$, in other words,
the dependence of the numerator and the denominator on $n_{\rm eff}$ should nearly cancel. 
This is indeed the case: if a sharply peaked redshift distribution
for the foreground galaxies around $z_{\rm f}$ is considered in (\ref{R_def}), one finds

\begin{equation}
R\simeq{3\over 2}\left({H_0\over c}\right)^2 {\Omega\over b} {g(w_{\rm f})
f_K(w_{\rm f}) p_{\rm f}(w_{\rm f})\over a(w_{\rm f}) \int dw~p_{\rm f}^2(w)}.
\end{equation}
Thus it is not surprising that $R$ does not depends on $n$ provided that the foreground
redshift distribution is narrow. To check this
assertion, three other power spectra have been tested: the Baugh \& Gazta$\tilde{\rm n}$aga (1996) power
spectrum (which behaves like $\propto k^{-2}$ at large $k$),

\begin{equation}
P(k)\propto{k\over \left[1+(k/k_{\rm c})^2\right]^{3/2}},
\end{equation}
where\footnote{$h=H_0/100$} $k_{\rm c}=0.05~h{\rm Mpc}^{-1}$, and two power law spectra $P(k)\propto k^n$, with $n=-1$ and $n=0$.
The values of $R$ from these
three power spectra are almost the same as the values plotted in Figure \ref{plot_R_val.ps}.
The difference is always smaller than 3\%, the worst case is for $n=0$, $z_{\rm f}=0.4$ for which
a variation of 5\% is obtained only for $\theta_{\rm c}=100'$. These results confirm that $R$ is
independent
of $n$ at least for $n\in [-3,0]$, in addition to be independent of scale. Hence, $R$ is
a direct estimator of the bias and the cosmology, nearly independent of the power spectrum and the
smoothing scale. Using other calculations of $Rb$ for different cosmologies, and
for $z_{\rm f}=0.4$, a power law fit to the $\Omega$ dependence of $Rb$ 
(with a zero cosmological constant) gives

\begin{equation}
R\propto {\Omega^{0.42}\over b}.
\end{equation}
The proportionality constant and the exponant $0.42$ should both depend on the foreground and
background redshift distributions.
Whether these numbers depend in a crucial way on these distributions will be
investigated in a future work, but it is worth to note that $Rb$ remains a discrimator of the
bias and the cosmology, almost independent of the power spectrum and the smoothing scale.

This encouraging result is motivating for the detection
of a possible variation of the bias with the scale, using the measurable quantity $\cal R$,

\begin{equation}
{\cal R}={R_{\theta_{\rm c}}\over R_{\theta_{\rm c}=1'}}={b(1')\over b(\theta_{\rm c})},
\label{cal_R_def}
\end{equation}
\begin{figure}
\centerline{
\psfig{figure=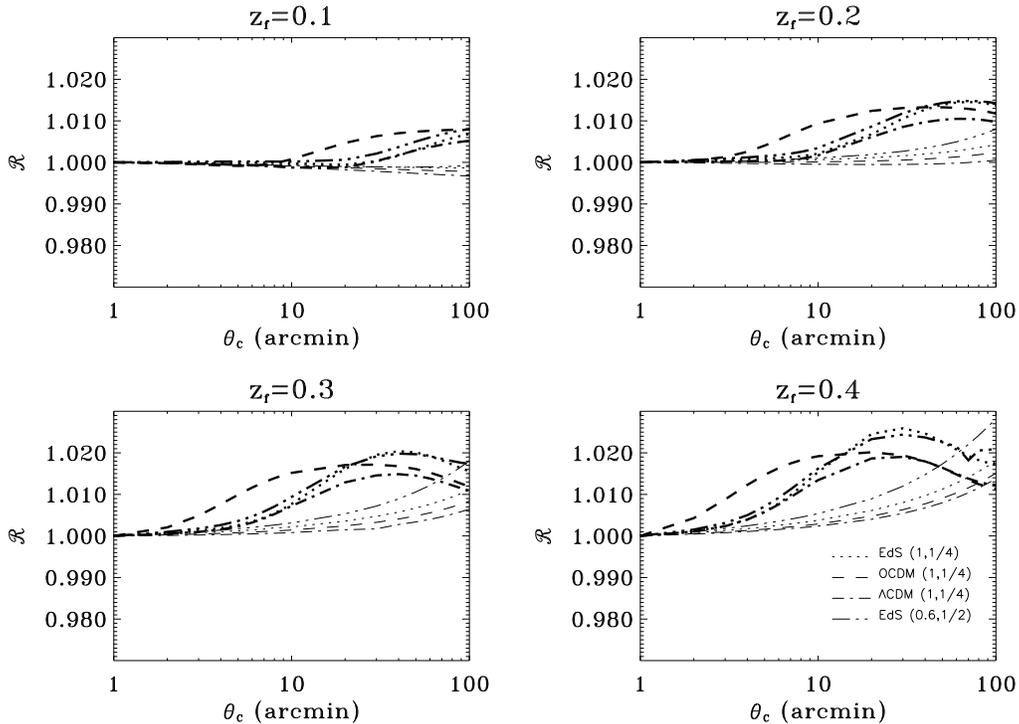,width=14cm}}
\caption{\label{plot_R.ps} The values of $\cal R$ (Eq. (\ref{cal_R_def})) versus the
smoothing scale are shown. The thick lines are for the non-linear power spectrum, while the
thin lines are
for the linear power spectrum. Whatever the cosmology, the largest variation expected for $\cal R$ is 
lower than about $2\%$, provided the bias parameter is constant with scale.}
\end{figure}
$\cal R$ has the advantage of being only bias dependent.
Figure \ref{plot_R.ps} shows $\cal R$ versus the smoothing scale
$\theta_{\rm c}$. For the wide range of scales $[1',100']$ considered here, this ratio remains very
close to $1$ with a variation smaller than $2\%$ whatever the cosmological model, and whatever the
redshift distribution of the foreground galaxies, provided that $b$ is scale independent. This is
the main result of this paper. It means that the
degeneracy between the bias parameter and the other cosmological parameters and the power spectrum
may be partially removed by choosing a particular filtering in the $\kg$ and redshift spaces.
The consequence here is to allow a measurement of the scale dependence of the bias without any
knowledge of the cosmological model.

According to Bardeen et al. (1986), the bias may be redshift dependent $b\propto 1+z$. The effect
of the introduction of this dependence has been calculated, and it changes the values of
$\cal R$ by less than 0.001\%, because
$p_{\rm f}(w)$ is chosen to be sufficiently narrow; only the amplitude of $R$ is changed, which
does not affect $\cal R$. Calculations with a different power spectrum (Baugh \& Gazta$\tilde
{\rm n}$aga
1996), and different redshift distributions for the background galaxies ($z_{\rm b}=0.7$ and
$z_{\rm b}=1.5$) have also
been performed, and it is found that the change of the shape of $\cal R$ versus $\theta_{\rm c}$ is
less than 0.5\% for
smoothing scales larger than $20'$, and insignificant for smaller scales. Thus $\cal R$ is an
estimator of the scale dependence of the bias almost independently of the redshift distribution
of the background galaxies and of any hypothesis of the
power spectrum and the cosmological parameters.

Next, estimates of the dispersion of $\cal R$ for a given number $N_{\theta_{\rm c}}$ of
fields of
size $\theta_{\rm c}$ are given. To calculate it, the dispersion of $M_{\rm ap}(\theta_{\rm c})
{\cal N}(\theta_{\rm c})$
and of ${\cal N}^2(\theta_{\rm c})$ are required. The former has already been calculated
(Eq.(\ref{S_N_def}) and Figure \ref{plot_S_N_NL.ps} give the signal-to-noise), but
the latter remains difficult to estimate. However, $\langle {\cal N}^2(\theta_{\rm c})\rangle $
can be assumed to be known very precisely from the existing and forthcoming large galaxy
catalogues\footnote{Like the CFRS, the LCRS, the 2dF, the SDSS, the VIRMOS survey, and the EIS
survey.}
because the redshift of the foreground galaxies is chosen to be quite low ($z\sim 0.3$). Therefore
the dispersion of $\langle {\cal N}^2(\theta_{\rm c})\rangle $ is neglected in estimating
the dispersion of $\cal R$.
If $\langle {\cal N}^2(\theta_{\rm c})\rangle $ is estimated from a different catalogue than the catalogue used
for $\langle M_{\rm ap}(\theta_{\rm c}){\cal N}(\theta_{\rm c})\rangle $, then it is important to have the
same selection criteria for the foreground galaxies of both catalogues. Otherwise, the bias factor
in the numerator and the denominator of (\ref{R_def}) may not be the same.
The dispersion of $\cal R$ will be derived in the limit of a small correlation between 
$M_{\rm ap}(\theta_{\rm c})$ and $ {\cal N}(\theta_{\rm c})$, which in fact
correspond to a determination
of the signal-to-noise of $\cal R$ rather than the complete calculation of the dispersion.
If one observes more than one field, the signal-to-noise of $M_{\rm ap}(\theta_{\rm c}){\cal N}(\theta_{\rm c})$ will
be reduced by a factor $\sqrt{N_{\theta_{\rm c}}}$, and if $N_{\theta_{\rm c}}\gg 1$, the
dispersion of $M_{\rm ap}(\theta_{\rm c}){\cal N}(\theta_{\rm c})/\langle M_{\rm ap}
(\theta_{\rm c}){\cal N}(\theta_{\rm c})\rangle$
is small compared to $1$ so that the dispersion of $\cal R$ is simply

\begin{equation}
\sigma_{\cal R}=\sqrt{{1\over N_{\theta_{\rm c}} {\cal S}^2_{\theta_{\rm c}}}+
{1\over N_{1'} {\cal S}^2_{1'}}}
\label{err_R}
\end{equation}
where ${\cal S}_{\theta_{\rm c}}$ is the signal-to-noise of
$M_{\rm ap}(\theta_{\rm c}){\cal N}(\theta_{\rm c})$ at the scale $\theta_{\rm c}$, given by 
Eq.(\ref{S_N_def}). $\sigma_{\cal R}$ is derived in Appendix A.

\begin{figure}
\centerline{
\psfig{figure=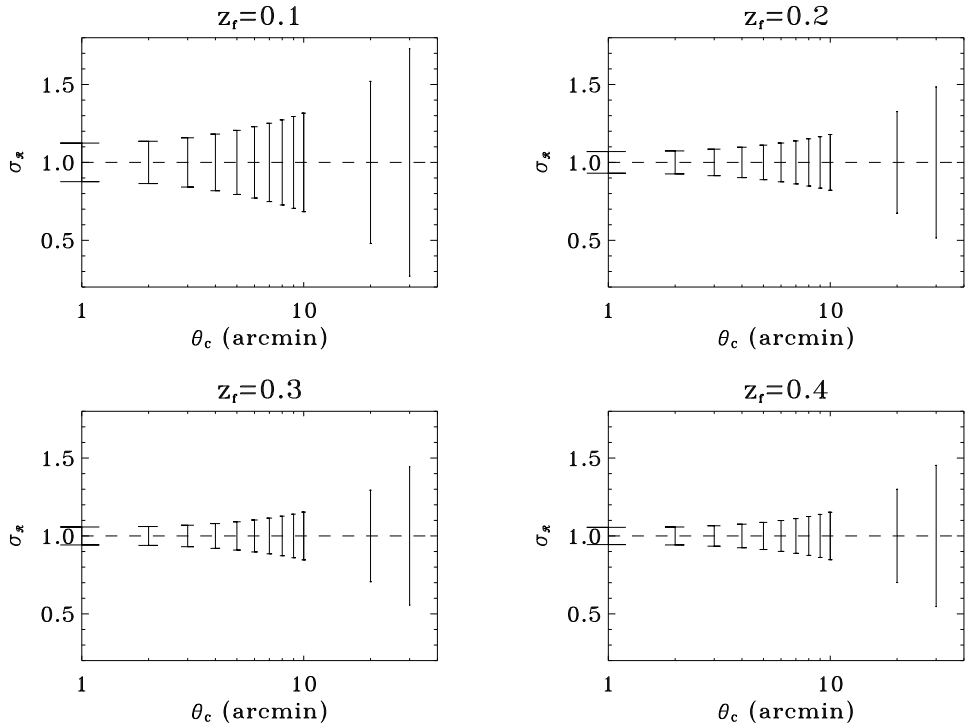,width=14cm}}
\caption{\label{plot_Er.ps} The dashed line show ${\cal R}=1$, and
error bars are the
standard deviation of $\cal R$ versus the
smoothing scale. It correspond to a $25$ square degrees field.}
\end{figure}

The reason why $\cal R$ is normalized to its value at $1'$ is to minimize the error on $\cal R$.
The dispersion $\sigma_{\cal R}$
is calculated by taking for ${\cal S}^2_{\theta_{\rm c}}$ and ${\cal S}^2_{1'}$
the values found in Figure \ref{plot_S_N_NL.ps}, averaged over the four cosmological models for each
smoothing scale, and for the non-linear power spectrum.
Figure \ref{plot_Er.ps} shows the expected error bars for a survey of $25$ square-degrees.
The number of fields of diameter $2\theta_{\rm c}$ (in arcmin) is given by $N_{\theta_{\rm c}}=
2.25\times 10^4/\theta_{\rm c}^2$ (the distance between two neighboring fields is choosen to
be $2\theta_{\rm c}$, see Appendix A for a discussion about the decorrelation properties
of two neighboring fields), and the
number densities of foreground and background galaxies are the same as above ($\bar N=5~{\rm gal}/
{\rm arcmin}^2$ and $n_{\rm b}=60~{\rm gal}/{\rm arcmin}^2$).
We see that a variation
of the bias of more than $20\%$ over the scale interval $[1',10']$ is detectable, but
if the foreground galaxies are located around $z_{\rm f}\simeq 0.1$, the error bars
are larger, which is
a consequence of the lower signal-to-noise in that case (see the upper left plot in Figure 
\ref{plot_S_N_NL.ps}).

A complete calculation of the cosmic variance of $\cal R$ remains to be done,
but the order of magnitude of the signal-to-noise calculated here shows the feasability of 
the measurement of the scale dependence of the bias parameter from surveys of a few ten square
degrees size.

%\begin{figure}
%\centerline{
%\psfig{figure=plot_R_Er.ps,width=9cm}}
%\caption{\label{plot_R_Er.ps} Standard deviation of $Rb$ versus the smoothing scale for
%four different cosmologies. The values
%of $Rb$ are those on Figure \ref{plot_R_val.ps} for the non-linear power spectrum
%(lower-right plot), for which $z_{\rm f}=0.4$ has
%been choosen. The error bars here are calculated assuming a survey of $25$ square degrees. High
%density universes are well separated from low-density universes, and $\Lambda$-dominated universes
%may be distinguished from low density universes.}
%\end{figure}

\section{Conclusion and discussion}

The cross-correlation between the aperture mass and number counts for a wide range of scales
$[1',100']$, and for different cosmological models has been calculated. The non-linear evolution
of the power spectrum has been taken into account.
It was found that the signal is dominated by this non-linear clustering.
At scales larger than about $8'$, the signal-to-noise ratio per field is $\sim
0.4$, close to the corresponding value calculated with the linear power spectrum.
At small scales,
the signal-to-noise per field decreases to $0.1-0.2$ at $1'$, while it is
close to zero for the linear power spectrum. Moreover, a significant additional decrease of the
signal-to-noise has
been found when the foreground galaxies are located at very low redshift ($\sim 0.1$).

The second part of this paper is focussed on the measurement of the scale dependence of
the bias, in the light of the previous results.
It has been shown that the quantity $Rb$, (where $R$ is defined as the cross-correlation of the
aperture mass-number counts
divided by the dispersion of the number counts), is almost independent of scale. The reason for this
is that the narrow filter applied in Fourier space and the assumed narrow redshift distribution
of the foreground galaxies reduce considerably the range of integration over the scales and the
redshifts in Eq. (\ref{MN_def}). This allows to approximate a general power spectrum as a
power law, for
which $Rb$ is independent on scale. Moreover, it has been shown that $Rb$ is almost
independent of the slope of this power law, thus $Rb$ is nearly independent of the shape of
a general power spectrum, in addition to be independent on scale. $R$ is
thus a direct measure of the bias times a function which involves the cosmological parameters,
and the redshift distributions of the foreground and background galaxies.
The quantity $\cal R$, which is
the ratio of $R$ at two different scales is only bias dependent; it is a measure of the ratio
of the bias at these two different scales. Even for a general cosmological model, $\cal R$ remains
almost independent of the power spectrum (because $R$ is also independent of it), the cosmological
parameters, the redshift distribution
of the foreground and background galaxies, if the bias parameter is independent of scale.
It has been shown that $\cal R$ may be used to
measure a pure scale dependence of the bias, and the expected signal-to-noise ratio for a survey of
$25$ square degrees size has been calculated.
A variation of the bias of more than 20\% over the scale range $[1',10']$ is detectable from this 
survey, provided that the variance $\langle {\cal N}^2(\theta)\rangle $
of the number counts is well known (from other
surveys for instance). Since the observer has the choice to select different redshift intervals
for the foreground galaxies,
it is in addition possible to detect a possible redshift dependence of the
variation of the bias with scale.

The method presented here allows a measure of a variation of the bias with scale if it
is larger than 2\% between $1'$ and $100'$. Below this limit, some residual couplings between the
cosmology, the power
spectrum, and the redshift distributions of the galaxies produce some variations in the
curve ${\cal R}(\theta_{\rm c})$.
However, for the largest scales considered here, it seems difficult to reach this limit in the near future
because a very large survey would be required. For example, if we want
a precision of $\pm ~0.2$ on $\cal R$ at $\theta_{\rm c}=100'$,
a survey of $900$ square degrees is required (with such a size, the error bars are smaller
than $0.02$ for scales below $10'$).

The foreground redshift distributions discussed here are available with photometric redshifts. For
the shear measurement, the
image quality must be good, and wide field cameras are necessary. A MEGACAM survey with 25
square degrees, UBVRI colors, and a median seeing of $0"6$ would be perfect.
%EIS could provide
%also some constraints, but the image quality is still uncertain. SDSS is ideal as regards
%the size of the field, but the image quality is probably not sufficient.

\section{Ackowledgements}

I am very grateful to P. Schneider, F. Bernardeau, Y. Mellier and  T. Erben for many stimulating 
discussions
and very useful suggestions. I thank S. Charlot who kindly calculated the precise redshift number
counts required for the estimation of the signal-to-noise. I am grateful to IAP for hospitality,
where part of this work has made some progress, and to the {\it programme nationnal de cosmologie}.
This work was supported by the "Sonderforschungsbereich 375-95
f\"ur Astro-Teilchenphysik" der Deutschen Forschungsgemeinschaft.

\section{References}

\parindent 0truemm
Bardeen, J.M., Bond, J.R., Kaiser, N., Szalay, A.S., 1986, ApJ, 304, 15

Baugh \& Gazta$\tilde{\rm n}$aga, 1996, MNRAS 280, L37

Bernardeau, F., van Waerbeke, L. \& Mellier, Y., 1997, A\&A 322, 1

Blandford, R.D., Saust, A.B., Brainerd, T.G. \& Villumsen, J.V.,1991, MNRAS 251, 600

Charlot, S., {\it private communication}

Fry, J.N. \& Gazta$\tilde{\rm n}$aga, E., ApJ, 1993, 413, 447

Jain, B \& Seljak, U. 1997 ApJ 484, 560

Kaiser, N., 1992, ApJ 388, 272

Kaiser, N., 1996, astro-ph/9610120

Kaiser, N., Squires, G., Fahlman, G. \& Woods, D., 1994, in: {\it
Clusters of Galaxies}, eds. 

{\hskip 1cm F. Durret, A. Mazure \& J. Tran
Thanh Van, Editions Frontieres}

Moesner, R., Jain, B., 1997, astro-ph/9709159

Peacock, J.A. \& Dodds, S.J., 1996, MNRAS 280, L19

Pell\'o, R., Miralles, J.M., Leborgne, J.F., Picat, J.P., Soucail, G., Bruzual, G., 1996, A\& A,
314, 73

Sanz, J.L., Martinez-Gonzales, E., Benitez, N., 1997, astro-ph/9706278

Schneider, P., 1996, MNRAS 283, 837

Schneider, P., 1997, astro-ph/9708269

Schneider, P., van Waerbeke, L., Jain, B., Kruse, G., 1997, astro-ph/9708143 (SvWJK)

Tyson, J. A., 1988, AJ, 96, 1

Villumsen, J., 1996a, MNRAS 281, 369

Villumsen, J., 1996b, MNRAS {\it submitted}

Villumsen, J., Freudling, W. \& da Costa, L., 1997, ApJ, 481, 578

\section{Appendix A: Evaluation of the standard deviation of $\cal R$}

In this Appendix, the standard deviation of $\cal R$ is derived in the limit of small
correlation between $M_{\rm ap}(\theta_{\rm c})$ and ${\cal N} (\theta_{\rm c})$, this gives
the signal-to-noise of the estimator $\cal R$.
$\cal R$ is defined in Eq.(\ref{cal_R_def}), which requires $R_{\theta_{\rm c}}$, defined in 
Eq.(\ref{R_first}). If a number $N_{\theta_{\rm c}}$ of fields of size $\theta_{\rm c}$ are
observed, the estimated value of $R$ is,

\begin{equation}
\tilde R={{1\over N_{\theta_{\rm c}}}{\displaystyle \sum_{i=1}^{N_{\theta_{\rm c}}}}\left[\tilde M_{\rm ap}(\theta_{\rm c})\tilde
 {\cal N} (\theta_{\rm c})\right]_i\over \langle {\cal N}^2(\theta_{\rm c})\rangle }.
\end{equation}
The tilde quantities always refers to the estimator of this quantity.
The observable $\tilde X_{\theta_{\rm c}}=
{1\over N_{\theta_{\rm c}}}{\displaystyle \sum_{i=1}^{N_{\theta_{\rm c}}}}\left[\tilde M_{\rm ap}(\theta_{\rm c})\tilde
 {\cal N} (\theta_{\rm c})\right]_i$ is introduced. It is assumed that its signal-to-noise is
simply ${\cal S}_{\theta_{\rm c}} \sqrt{N_{\theta_{\rm c}}}$, where ${\cal S}_{\theta_{\rm c}}$ is
the signal-to-noise of $\tilde M_{\rm ap}(\theta_{\rm c})\tilde {\cal N} (\theta_{\rm c})$ in
one field of size $\theta_{\rm c}$ (see Eq.(\ref{S_N_def})). 
This assumption implicitly neglect the correlation between two
neighboring fields, in other words, terms like $\langle \left[\tilde M_{\rm ap}(\theta_{\rm c})\tilde
 {\cal N} (\theta_{\rm c})\right]_i\left[\tilde M_{\rm ap}(\theta_{\rm c})\tilde
 {\cal N} (\theta_{\rm c})\right]_{j\ne i}\rangle$ are neglected. This point is discussed at the end of this Appendix. Thus it is possible
to express $\tilde X_{\theta_{\rm c}}$ as a function of the ensemble average $\langle X_{\theta_{\rm
 c}}\rangle$, and a random variable $E_{\theta_{\rm c}}$,

\begin{equation}
\tilde X_{\theta_{\rm c}}=\langle X_{\theta_{\rm c}}\rangle(1+E_{\theta_{\rm c}})
\end{equation}
where $\langle X_{\theta_{\rm c}}\rangle=\langle M_{\rm ap}(\theta_{\rm c}) {\cal N} (\theta_{\rm c})\rangle$, and
$E_{\theta_{\rm c}}$ is a random variable such that $\langle E_{\theta_{\rm c}}\rangle=0$.
The limit of small correlation $\langle X_{\theta_{\rm c}}\rangle\simeq 0$ is used, then the
dispersion of $E_{\theta_{\rm c}}$ is $\langle E^2_{\theta_{\rm c}}\rangle=
1/(N_{\theta_{\rm c}}{\cal S}^2_{\theta_{\rm c}})$.

$\tilde{\cal R}$ was defined as

\begin{equation}
\tilde{\cal R}={\tilde X_{\theta_{\rm c}}\over \langle {\cal N}^2(\theta_{\rm c})\rangle}
{\langle {\cal N}^2(1')\rangle\over \tilde X_{1'}}.
\label{R_estim_def}
\end{equation}
We want to calculate the dispersion $\sigma_{\tilde {\cal R}}^2=\langle \tilde{\cal R}^2\rangle 
-\langle \tilde{\cal R}
\rangle^2$. As explained in Section 4, the dispersion of the measurement of ${\cal N}^2$
is neglected, this is why ensemble average values $\langle {\cal N}^2\rangle$ comes directly in (\ref{R_estim_def}).
Provided that the number of fields is large, $N_{\theta_{\rm c}}\gg 1$, the dispersion
$\langle E^2_{\theta_{\rm c}}\rangle$ is small compare to $1$, thus the calculation
of $\sigma_{\tilde {\cal R}}^2$ may be restricted to the second order in $E_{\theta_{\rm c}}$.
In addition, neglecting the cross-correlation terms like $\langle E_{\theta_{\rm c}} E_{1'}\rangle$
(this is discussed at the end of this Appendix),
the dispersion of $\tilde{\cal R}$ is given by,

\begin{equation}
\langle \tilde{\cal R}^2\rangle \simeq {\langle {\cal N}^2(1')\rangle^2\over \langle {\cal N}^2
(\theta_{\rm c})\rangle^2}  {\langle X_{\theta_{\rm c}}\rangle^2\over
\langle X_{1'}\rangle^2} \langle1+E^2_{\theta_{\rm c}}+2E_{\theta_{\rm c}}-2E_{1'}+3E^2_{1'}+...\rangle
=(1+\langle E^2_{\theta_{\rm c}}\rangle +3\langle E^2_{1'}\rangle+...),
\end{equation}
and,

\begin{equation}
\langle \tilde{\cal R}\rangle^2 \simeq {\langle {\cal N}^2(1')\rangle^2\over \langle {\cal N}^2
(\theta_{\rm c})\rangle^2}  {\langle X_{\theta_{\rm c}}\rangle^2\over
\langle X_{1'}\rangle^2} \langle 1+E_{\theta_{\rm c}}-E_{1'}+E^2_{1'}+...\rangle^2=1+2 \langle E^2_{1'}\rangle+...
\end{equation}
and finally,

\begin{equation}
%\langle \tilde{\cal R}^2\rangle-\langle \tilde{\cal R}
%\rangle^2
\sigma^2_{\cal R}=\langle E^2_{\theta_{\rm c}}\rangle +\langle E^2_{1'}\rangle=
{1\over N_{\theta_{\rm c}} {\cal S}^2_{\theta_{\rm c}}}+{1\over N_{1'} {\cal S}^2_{1'}},
\label{calc_err_R}
\end{equation}
and the signal-to-noise of $\cal R$ is $1/\sigma_{\cal R}$.
The derivation of (\ref{calc_err_R}) implies some simplifications mentioned above,
which have to be discussed.
From a practical point of view, the $N_{\theta_{\rm c}}$ fields should be extracted from one
large survey (or at least, the largest scale of interest should be the smallest scale of a sub-field
from a fragmented survey). Thus we expect that two neighboring fields are correlated and that two
fields centered
on the same position but with two different smoothing scales are also correlated.
It has been shown in SvWJK (Figure 8) that
the use of compensated filters allows to neglect these correlations\footnote{Even if the
compensated filter
used here is not excatly the same as in SvWJK, the decorrelation properties of this type of filter
are the same.}; the aperture mass $M_{\rm ap}(\theta_{\rm c})$ of two neighboring fields of radius
$\theta_{\rm c}$ with a
separation $\theta_{\rm c}$ between the two centers
are decorrelated by a factor of 10, and more than a factor 100 if the separation is 
$2~\theta_{\rm c}$. The aperture mass of two coincide fields with a factor
of 5 between the two smoothing scales are also decorrelated by a factor of 10. Moreover, if these
two fields are off-centered (which can be done in practice), the decorrelation is stronger.
For example, two fields with a scale ratio of 2 are decorelated by a factor larger than 10 if
they are off-centered by only half of the radius of the largest field.
These decorrelation properties, direct consequences of the use of compensated filters,
justify the assumptions made above to derive (\ref{calc_err_R}).

\end{document}